\documentclass{article}

\usepackage{graphicx}
\usepackage{amssymb}
\usepackage{amsmath}

\begin{document}

\title{Observables and Cohomology Classes for Classical Gravitational Field}

\author{M. Iftime \thanks{Boston, Massachusetts, Email:miftime@gmail.com}}

\date{}
\maketitle

\begin{abstract}

Some of the most outstanding questions in the field of gravitation and geometry remain unsolved as a result of our limited understanding of the global structure of the spacetime geometry and the role played by global spacetime diffeomorphism group in quantum gravity. Some insight into these important questions may be gained by looking at certain aspects of general covariance and observables in classical gravitational theory. 

I this paper I shall define as set of classical geometric observables of the gravitational field by which I mean $Diff(M)$-gauge invariant cohomology classes defined on a Lorentzian structure. They represent global characteristics of the physical gravitational phenomena, are linked to the topology of the spacetime, and can be used in constructing new Lagrangians. 

The problem of finding a complete set of data out of observables is related perhaps to the fact that at present moment, manifolds in dimension $4$ and above cannot be effectively classified. One could interpret this result as a pointer to the possibility that there might be spacetimes with different topologies (i.e., different global characteristics) which have indistinguishable local spacetime geometry.

\end{abstract}

\date{}

\maketitle

\pagebreak

\tableofcontents

\pagebreak

\section{Introduction}

\subsection{Observables}

The laws of physics that determine classical dynamics of gravitational field is well understood. Spacetimes are the arenas in which all physical events take place; an event is a point in spacetime specified by its time and place.  
In relativistic context, a spacetime is described by a differentiable manifold $M$ of dimension $4$, and a gravitational metric field $g$. 

General relativity is background-independent theory, defined on a manifold $M$ endowed with no {\em a priori} background metric structure. The metric $g$ is the sole dynamical variable \footnote{Apart from the value of the Newtonian coupling constant which is measurable in laboratory experiments.} that is a solution of Einstein's equations. We implicitly assume that the gravitational metric $g$ is defined on the entire manifold $M$, and that all points in $M$ are regular.

Observables in general relativity are analogous gauge-invariant quantities in classical electromagnetic theory. The gauge transformations are spacetime diffeomorphism. Einstein's field equations take the same form in any coordinate system, and they are invariant under spacetime diffeomorphisms. 

Physicists use two interpretations of the concept of diffeomorphism-invariance: "passive" and "active".\cite{Rovelli} Passive diffeomorphism-invariance refers to invariance under the pseudo-group\footnote{See \cite{Olver} for recent investigations of $D(M)$ and a direct approach to the Cartan structure equations for pseudo-groups} $D(M)$ of local difeomorphisms on $M$. It is clear that any theory can be made invariant under passive diffeomorphisms, because coordinates do not have a physical meaning. A spacetime is independent of any observer, however in describing physical phenomena (which occur at certain moments of time in a given region of space), each observer chooses a convenient coordinate system. 

Global diffeomorphisms appear as the active transformations of $M$.
The group $Diff (M)$ \footnote{One can restrict to global diffeomorphisms with compact support, due to the behaviour at the asymptotic regions of $M$\cite{Isham}. See \cite{IftimeStachel} for a discussion on the hole argument} is the genuine gauge group of general relativity.  Invariance under $Diff(M)$ is a key feature of Einstein metric theory of gravity, called general covariance. This means that if $g$ is a solution of Einstein's equations, then for any $f\in Diff(M)$, the pullback metrics $f^{*}g $ are also solutions, and moreover, $g$ and $f^{*}g $ represent the same physical gravitational field.  

A consequence of the invariance of the physics laws under coordinate tranformations implies the unobservability of the spacetime coordinates by which we identify events in spacetime. Moreover, the gravitational metric tensor $g_{ab}$ \footnote{The gravitational metric field represents not only the chrono-geometrical structure of space-time, but also the potentials for the inertio-gravitational field} defined as a function of spacetime coordinates is also not observable. 

Invariance under $Diff(M)$ has the consequence that spacetime points (and the spacetime fields) have no direct physical significance, as emphasized by Einstein\cite{Einstein}:

\begin{quote}
"All our spacetime verifications invariably amount to a determination of spacetime coincidences. If, for example, events consisted merely in the motion of material points, then ultimately nothing would be observable but the meeting of two or more of these points."
\end{quote}

The only things that have immediate physical meaning are the gauge invariant objects. 
Two space-time solutions of Einstein's equations are physically equivalent if one is isometric to the other, so the only thing that has a direct physical significance is the superspace $Geom(M): = Metrics(M)/Diff(M)$ of isometry classes of  metrics on $M$. A physical space-time can be identified with a point in the superspace. \footnote{$Geom(M)$ doesn't have a manifold structure, but in most cases is a stratified manifold, with each strata consisting of space-times with conjugate isometry groups.\cite{Palais},\cite{Fisher}.}

We shall refer to $Diff(M)$-invariant objects as {\em true} observables, and the $D(M)$-invariant ones  as {\em local} observables.
Observables are uniquely determined by the physical situation, and given a complete set of true observables would completely characterize the gravitational field under consideration, by removing the ambiguity created by general covariance.\\

{\em Examples:}\\

The situation in general relativity is quite different from most other field theories, where the structure of spacetime is assumed in advance, and the gauge transformations occur at a fixed spacetime points.

The attempt to localize or individuate\cite{Stachel} the spacetime points (events), will always depend on the choice of a metric in the neighborhood of the point (i.e., local gauge in the bundle language). If the metric is generic (i.e., solutions of Einstein field equations with no symmetry group, or in other words, spacetime manifolds with no Killing vectors) Komar \cite{Komar} constructed a set of "intrinsec" spacetime coordinates to identify the spacetime points uniquely. Komar's coordinates consists of a set of four functionally independent invariant scalars (local scalar invariants constructed from the metric and its derivatives must be functions of the metric and the curvature. On $M$ at most $4$ such scalars can be functionally independent), which are defined uniquely in terms of the local geometry of the spacetime; they are local observables, and don't depend on asymptotic boundary conditions.

If the metric has symmetries, then some further choice of parameters on the orbits of the symmetry group is needed. To actually measure the components of the Riemann tensor in order to compute the invariants may ultimately involve the determination of spacetime coincidences, involving material points, or perhaps better, material detectors of finite size operating over a finite time interval, so that we can only measure average values of the components over a finite space-time region. \footnote{This is what Bohr and Rosenfeld found to be the case for the components of the electromagnetic field.}

In practice one can define local observables with respect to matter or light degrees of freedom.  The basic measurements are the distance between two spacetime events which consist of collision of material points or light quanta (in a physical sense it depends on the existence and distinction of some matter fields). For empty spacetime regions, one often considers 'test matter'. The stress-energy tensor of the matter is so small that its effect on the gravitational field may be neglected. For a particle, this means that its mass is so small that its effect may be neglected. One could say that the body has a "passive", but not "active" stress-energy tensor, so that the effect of the gravitational field on the body can be detected, but not the effect of the body on the gravitational field.

\subsection{Functor Bundles}

(Gauge-)natural functor bundles provide a unified approach to all 
classical field theories.

{\em Examples:}

Electromagnetic theory is a gauge theory of principal connections on a principal bundle $(P\to M;G )$ with the gauge group $G=U(1)$ acting at fixed points $p$ in $M$. The values of the gauge field depends locally on spacetime points.  However one can model electromagnetic theory on a gauge-natural bundle as follows: (see e.g., \cite{IftimeStachel}). Being $G$-equivariant, principal connections on $P$ (the gauge potentials), can be identified with global cross-sections of a gauge natural bundle functor -- the associated  bundle $E = J^{1}P/G$, where $J^{1}P$ denotes the first order jet manifold of sections of $P$. The role played by the configuration (structure) bundle $(P\to M)$ is that it selects a certain class of global cross-sections of $E = J^{1}P/G$.  Maxwell's electromagnetic theory is a rule for selecting theclass of cross-sections of $P$,  1-form fields that obey
the linear, gauge-invariant field equations derived from the Maxwell lagrangian. While Born-Infeld electrodynamic theory is a rule for selecting a (different) class of cross-sections of $P$, 1-form fields that obey the non-linear (but gauge-invariant) field equations derived from the Born-Infeld lagrangian.\\

General relativity can be interpreted both as a natural and gauge natural theory.\cite{Iftime2008} The relevant group of general relativity is $G = SO(3,1)$, or more precisely the (restricted\footnote{Some authors refer to the full Lorentz group $SO(3,1)$ when they actually mean the restricted Lorentz group $SO^{+}(3,1)$, which is the set of Lorentz transformations preserving both orientation and the direction of time, which is the identity connected component of $SO(1, 3)$} ) Lorentz group $SO^{+}(3,1)$.

A gravitational field has a 'natural' geometric realization as a $G$-structure of first order on $M$. By a $G$-structure we mean a principal bundle $(P\to M; G )$, a reduced \footnote{i.e.,
$G$ is a reductive group of $Gl(4;\mathbb{R})$, because one can write $sl(4) = so(1,3)\oplus g'$, where $g' =\{X\in sl(4),\:X^{t}= X\}$}
principal subbundle of the linear frame bundle $FM$. "Naturality" means that the group $Diff(M)$ acts on $P\in FM$ such that spacetime diffeomorphisms lift uniquely to
an affine transformations of the basis vectors at each point of $P$.
The fact that the principal bundle $P$ consists of frames is considered part of the data. In general relativity, principal bundle isomorphisms are "soldered" to the spacetime diffeomorphisms. The cannonical soldering form is what "ties" the underlying (principal) bundle $P$ to the local geometry of the spacetime manifold $M$. Two Lorentzian structures $P$ and $P'$ on $M$ are said to be equivalent if and only if the associated Lorentzian metrics $g$ and $g'$ are isometric.

A 'gauge-natural' geometric realization of a gravitational field can be given by a global cross-section $\sigma$ of the associated fiber bundle $(F^{*}M/G\stackrel{p}{\longrightarrow} M)$ of $G$-related coframes on $M$. We shall conveniently represent a coframe at $p\in M$ as the 1-jet $u_{p}=j^{1}\phi $ of a local (chart) diffeomorphism  $\phi: U_{p}\to \mathbb{R}^4$ such that $\phi(p)=0$. 

The group $Diff (M)$ of global spacetime diffeomorphism acts on the space of the cross-sections $\Gamma (F^{*}M/G)$ as follows: $f^{*}(\sigma): = f^{*}\circ \sigma \circ f^{-1}$, where $f^{*}: F^{*}M\to F^{*}M$ is the induced isomorphism on the bundle of coframes. This implies, that for any $f\in Diff(M)$ there is an isomorphism $(f^{*}, f): F^{*}M\to F^{*}M$ defined by 
$f^{*}(j^{1}_{p}\phi) = j^{1}_{f(p)}(\phi\circ f^{-1}\mid _{U_{p}})$ such that $f^{*}(P)=P$.

It is clear that the Lorentzian structures on $M$ are in bijective correspondence with the global cross-sections of $F^{*}M/G$ via the relation $P=p^{-1}(\sigma(M))$, and two Lorenzian structures are equivalent iff the corresponding cross-sections are $Diff(M)$-related.

The following sequence of maps summarizes the three geometric representations of a gravitational field:
$$Metrics(M)\stackrel{\Pi}{\longrightarrow} Geom(M)\stackrel{\varphi }{\cong}\mathcal{M}_{SO(3,1)}\stackrel{\phi 
}{\cong}(\Gamma(F^{*}M/G))/Diff(M)$$

The images of a Lorentzian metric $g$ represents the (unique) physical gravitational field defined by $g$ defined by a point $\Pi(g)$ in the superspace $Geom(M)$ (metric approach), an isomorphism class $\varphi(\Pi(g))$ in the  moduli space $\mathcal{M}_{SO(1,3)}$ of all isomorphism classes of Lorentzian structures on $M$ (natural bundle approach), or a class $\phi(\varphi(\Pi(g)))$ of $Diff(M)$-related global cross-section of the gauge-natural bundle $F^{*}M/G\to M$ (gauge-natural bundle approach).

By using this functorial bundle approach in general relativity, one can gain a number of other advantages. For example, the difference between the gauge transformations in general relativity and  electromagnetic theory is clearer in this approach. Functors are useful concepts in keeping track of the relationships between local and global data, in the sense that Lorentzian $G$-structures can be naturally defined as (local) gauge fields. A global cross-section of $(F^{*}M/G\to M)$ can be represented by a family of local cross-sections of $(P\to M)$ related via local gauge transformations. It is easy to work with coframes in the process of jet prolongations for $G$-structures used to construct true observables for the gravitational field and to define local-geodesic normal spacetime coordinates systems.\cite{Coleman}

\section{Construction of true observables for classical gravitational field}
 
Let $M$ and $G=SO(3,1)$ be fixed. By a characteristic class of the gravitational field we mean a way of associating to any Lorentzian structure $P$ an element $c(P)$ in some cohomology group $H^{*}$, such that for any $f\in Diff(M)$ then $c(f^{*}P) = f^{*}(c(P))$. (on the left is the class of the pullback $f^{*}P$, while on the right is the image of the class $c(P)$ under the induced map in cohomology)

From definition implies immediately that $c(P)\in H^{*}$ is an invariant of the isomorphism class of $P$ in $\mathcal{M}_{SO(1,3)}$, and  that therefore $Diff(M)$-invariant quantities which define true physical observables of the gravitational fields. In category language, $c$ can be viewed as a natural transformation from the set $\mathcal{M}(SO(3,1)$ to a cohomology functor $H^*$.

We shall define two types of true observables for the gravitational field, depending on the cohomology group they are defined.

\section{$\delta$-cohomology classes for gravitational field}

Let $P$ be a Lorentzian $G$-structure that represents the gravitational field under consideration. In detail, by Lorentzian structure we shall mean a triplet 
($(P\stackrel{\pi}{\longrightarrow}M),\rho,\theta$), where
$G = SO(3,1)$, $\pi: P\to M$  is a principal reduced subbundle of the linear frame bundle $FM$, $r:G\to GL(4, \mathbb{R})$ a linear represention of the structure group $G$, and $\theta:TP\to\mathbb{R}^4$ the soldering\footnote{The soldering form is
the unique 1-form on $P$ that "cancels" a pullback. It is defined cannonically by $ \theta_p(\xi) = p^{-1}d\pi(\xi)$, where $\xi$ is a tangent vector to $P$ at $p$ and $d\pi$ is the differential of the projection map.} 1-form $\theta\in \Lambda^{1}_{horiz}(P;\mathbb{R}^4)^{G}$ (which means $\theta$ is $G$-invariant (i.e., $R_a^{*}\theta = a^{-1}\theta $, where $R_a$ is right translation by $a\in G$), strongly horizontal(i.e., vanishes on vectors tangent to the fibers $\theta(X)=0$ for $X$ if $\pi_{*}X = 0$).

\subsection{The structure form of a Lorentzian geometry, and $\delta $- cohomology}

Let $V = \mathbb{R}^4$ and $V^*$ its dual. Let $L^{k}(V)=\otimes ^{k}V^{*}$ the space of multilinear forms on $V$, and $L^{k}(V)^{G}$ the subspace of $G$-invariant forms. Let $S^{k}(V^{*})$ and $\Lambda^{k}(V^{*})$ the subspace of symmetric, respectively skew-symmetric covariant forms on $V$. 

Let $S(V^{*})=\oplus^{k}S^{k}(V^{*})$ the symmetric algebra (which is in effect the same as the ring of real-valued polynomials on $\mathbb{R}^4$), and $\Lambda(V^{*})=\oplus^{k}\Lambda^{k}(V^{*})$ the exterior algebra on $V$.

The bigraded commutative algebra $V\otimes S(V^{*})\otimes\Lambda(V^{*})$ together with the 'antisymmetrization' linear maps: $\delta_{k,l} : V\otimes S^{k}(V^{*})\otimes\Lambda^{l}(V^{*})\longrightarrow  V\otimes S^{k-1}(V^{*})\otimes \Lambda^{l+1}(V^{*})$ satisfy $\delta^{2} = \delta_{k-1, l+1}\circ \delta_{k,l} = 0$.

The Lorentz group is the matrix Lie subgroup in $GL(4, \mathbb{R})$ which preserves the quadratic (Minkowski) form $\eta = diag(1,1,1,-1) $. Its Lie algebra  $\frak g = \frak {so}(1,3)$ is a subspace of $\frak {gl}(4)= Hom(V,V)= V\otimes V^{*}$ consisting of the skew-symmetric matrices. The first prolongation of $\frak {g}$ is defined as the subspace $\frak g^{1}$ of symmetric elements in $Hom(V, \frak {g})$, and by induction for $k > 0$, the $k$-th order prolongation is defined by $\frak{g}^k = [V\otimes S^{k+1}(V^{*})]\cap[\frak{g}\otimes S^{k}(V^{*})] $.

For $k, l >0$, denote $C^{k,l} = \frak{g}^{k-1}\otimes\Lambda^{l}(V^{*})$ (where $\frak{g}^{-1} = V$, $\frak{g}^{0} =\frak g$). The restrictions of $\delta$-maps to $C^{k,l}$ defines a cochain $\{C, \delta\}$. The corresponding cohomology groups,  $H^{k,l}(M,\frak{g})$, are the $\delta $-cohomology groups.\footnote{The only non-zero groups are of the form $H^{k,2}$, and they contain the obstructions from integrability (local flatness) for the Lorentzian structures.}\cite{Spencer, Sternberg}

One can define the structure form (see e.g., \cite{Fujimoto, Kobayashi}) of the Lorentzian geometry as follows: to each horizontal subspace  $H$ at $u_{p}\in P$\footnote{$H$ exists, but not unique until a connection on $P$ is specified}
one can define a horizontal, $G$-invariant form on $P$ one can associate a form:

$c_{H}(P)\in Hom(V\wedge V, V)$ defined by $c_{H}(P)(X^{H}_{1},X^{H}_{2}) = d\theta (\theta (X^{H}_{1}), \theta (X^{H}_{2}))$ where $H\stackrel{\theta }{\cong}V$. One can check that its cohomology class $c(P)$ in $H^{0,2}$ is independent of the choice of $H$, and for any isomorphism $f\in Diff(M)$ then $c(f^{*}P) = f^{*}c^{0}(P)$. It implies that $c(P)$ is an invariant of the isomorphism class of the Lorentzian structure $P$, and so it represents a true observables of the physical gravitation field determined by $P$.

Intuitively $c(P)\in H^{0,2}(M,\frak{g})$ can be interpreted as the part of the torsion of the Lorentzian structure $P$ \footnote{On a $G$-structure $P$ one can define many different principal connections, which in turn can have different torsions. In spite of this, there is an independent notion of torsion of $P$.} that is independent of the choice of the soldering form $\theta$. This structure forms were first time introduced by Cartan\cite{Cartan} as the "aparent torsion" in his algorithm for differential systems. 

Since $\delta_{1,1}$-map is an isomorphism implies that for any $G$-invariant tensor $t\in \frak{g}^{0}\otimes \Lambda^{2}(M)$ there is a unique $G$-connection having $t$ as the torsion tensor. Let  $\Gamma $ a principal connectionon on $P$ with  torsion tensor $\tau_{\Gamma }(z)$ , then $c(P) = p\circ \tau_{\Gamma }$.

Since $H^{0,2} = 0$ implies that $c(P) =0$, and so the torsion tensor $\tau_{\Gamma }(z)$ is "cohomologous to zero" in the cohomology group $H^{2,0}$. Intuitively, $c^{0} = 0$ gives the "integrality condition" for the existence and uniquence of the torsionfree metric (Levi-Civita) connection \footnote{The inertial-gravitational potential of the physical gravitational field}. There is an analogy with the situation in the electromagnetic theory where the electromagnetic field $F$ is cohomologous to zero in the DeRham group $H^{2}(M)$ implies the existence of the  electromagnetic potential 1-form $A$ such that $dA =F$.

\subsection{The process of prolongation}

A matrix in the general linear group $GL(V)$ can be representated as an Jacobian matrix at $0$ of a diffeomorphism $\varphi: R^{4}\to R^4$ with $\varphi (0) = 0$.

Let $k > 0$.  The $k$th order linear group $GL^{k}(4) = \{j^{k}\varphi,\,\varphi\in D(\mathbb{R}^{4}), \, \varphi(0)= 0\}$.
describes intuitively how a $k$th order Taylor polynomial transformsms under changes of coordinates (or equivalently, diffeomorphisms).

The $k$th order prolongation of the Lorentz group $G$ is defined as a closed subgroup $G^k$ of $GL^{k}(4)$ consisting of $k$th jets $j^{k}_{0}\varphi $ such that the Jacobian matrices $a^{j}_{i} = (x^{i}\circ \varphi_{,i})$ preserve the Minkowski product $\eta$ (i.e.,  $\eta_{ij}a^{i}_{k}a^{j}_{l} = \eta_{kl} $).

Since $J^{k}_{0}(V,V)$ can be identified with the space of polynomial maps on $\mathbb R ^4$ of degree less then or equal to $k$, the Lie algebra of $GL^{k}(V)$ has the representation  $\frak {gl}^{k}(V) =\displaystyle \oplus _{i=1}^{k} (V\otimes S^{i}(V*)$, where $S^{i}(V^{*})$ is isomorphic to the set of homogeneous polynomial of degree $i$ on $\mathbb R ^4$. The Lie algebra of $G^k$ can be written as $\frak{\tilde g}^{k} = \oplus_{i=0}^{k-1} \frak {g}^{i}$ 
forms a graded Lie algebra (with the induced bracket operation from the spaces $V\otimes S^{i}(V*)$), a Lie subalgebra of $\frak {gl}^{k}(V)$.\cite{KMS}

In order to construct true observables, we shall now use the gauge-natual geometric interpretation of the gravitational field and work with coframes.

Let $\sigma\in \Gamma (F^{*}M/G)$ representing the Lorentzian structure $P$.  Such cross-section $\sigma$ can be specified by giving at each point $p\in M$ an equivalence class of $G$-related coframes  $u_{p}=j^{1}\phi $, where $\phi: U_{p}\to \mathbb{R}^4$ is a local diffeomorphism such that $\phi(p)=0$.

For $k=1$, $F^{1}M = FM$ and $P^{1}= P$. For $k=2$, $P^{2}$ denotes the first-order jet prolongation of $P$. $P^{2}$ represents a second-order structure on $M$ with the structural group $G^{1} = G\times\{0\}$.\cite{KMS}

$P^{2}$ is an affine structure on $M$, consisting of the field of equivalence classes of $G^{1}$- related $2$-coframes $u^{2}_{p} =j^{2}\phi $, where $\phi:U_{p}\subset M\to V$, $\phi )p)= 0$, $p\in M$.

The previous condition $c(P) = 0$ implies that the Lorentzian metric affine structure of $P^2$ is completely determined by $P$. So, further prolongations of $P$ can be obtained by prolonging the metric affine structure $P^2$.
 
For $k > 1$, the $k$th prolongation $P^{k}$ of $P$ can be similarly defined by giving a global cross-section of the associated fiber bundle $F^{*k}M/G^{k}$ ( where $F^{*k}M = J^{k}(F^{*}M) $ denotes the bundle of $k$th coframes) which specifies at each $p\in M$ a field of classes of $G^k$- related $k$-coframes.

One can define a generalized soldering form $\theta = (\theta^{0},\cdots,\theta^{k-1})$ on $P^k$ as a $V + \frak{g} +\cdots +\frak{g}^{k-1}$ - valued horizontal 1-form defined as $c_{H}\in Hom(V\wedge V, V + \frak{g} +\cdots +\frak{g}^{k-1})$, 
$c_{H} (X_{1}, X_{2}) = d\theta (\theta (X^{H}_{1}), \theta (X^{H}_{2}))$,
where $H\cong V$ is a horizontal space at a point $u_{p}\in P^{k}$.

One can check\cite{Ogiue} that cohomology classes $c^i\in H^{i, 2}(M;G)$ of 
the $Hom(V\wedge V,\frak{g}^{i-1})$- components of $c_{H}$ are independent of the choice of the horizontal space $H$. (where $ \frak{g}^{-1}\cong V$, $\frak{g}^{0}=\frak{g}$) 

If $f\in Diff(M)$ is an isomorpfism of $P$  then $c((f^{k})^{*}P^{k}) = ((f^{k})^{*}c)P^{k}$. 
It is clear that the group $Diff(M)$ induces a natural action of the groupoid $J^{k}(M,M)$ of the $k$th jets of spacetime diffeomorphisms on the space of $k$th order prolongations of Lorentzian structures $J^{k}(F^{*}M/G)$: 
any isomorphism $f\in Diff(M)$ of $P$ induces an isomorphism of the corresponding $k$th prolongations 
$(f^{k},f)$ of $P^{k}$ defined as a bundle morphism $f^{k}(j^{k}_{p}\varphi) = j^{k}_{f(p)}(\varphi\circ f^{-1}\mid _{U_{p}})$ of  $F^{k*}M$ such that $f^{k}(P^{k})=P^{k}$. \footnote{The definition of $f^{k}$ is independent of the choice of the representative class of local diffeomorphism $\varphi $ at each $p\in M$.}

It implies that $c$, the structure tensor of $P^k$, is actually a $Diff(M)$- invariant of the isomorphism class of the Lorentzian structure $P$, and therefore represents a true observable for the gravitation field determined by $P$.

The $H^{0, 2}$- component of $c$ is exactly the the structure form of $P$, as defined previously. So $c^{0} = 0$ .
The component $c^{1}\in H^{1, 2}(\frak{g})$ is related to the curvature form on $P$(is non-zero for generic spacetimes). 

For $ i >1$,  $H^{i, 2} = 0$, which implies that all $i > 1$-components of $c$ are all zero.  This is in agreement with the intuitive analysis of the prolongation procedure for Einstein's equations \cite{Olver} when we get a family of derived differential equations from the original Einstein's equations with differentiating with respect to the spacetime (independent) variables. 
\footnote{Einstein's equations can be represented as a variety $S$ in $J^{2}(F^{*}M/G)$, i.e., a rule of global cross-sections $\sigma$ such that $j^{2}\sigma$ takes value in $S$.}

\subsection{Local characterization of the metric}

The prolongation projections $p^{k}_{k-1}:P^{k}\to P^{k-1}$ determine an affine connection $\Gamma ^{k-1}$ on $P^{k-1}$. At each prolongation step $k$, a $k$-coframe $j^{k}_{p}\varphi $ defines (1) by truncation a $(k-1)$-coframe $j^{k-1}_{p}\varphi $ ( which is (gauge-related) to the $(k-1)$-coframe at $p$ in $P^{k-1}$), and (2) a first order Taylor approximation of a $k-1$ jet at nearby points $q\in U_{p}$ in the coordinate difference $x(q)-x(p)$. The composition of the two maps (1) and (2) results in a invertible map $\Pi_{p,q}:F^{(k-1)*}M\to F^{(k-1)*}M$ which is Lorentz invariant, therefore it defines a parallel transport on $P^{k-1}$. (see e.g., \cite{Coleman}) If $\Gamma ^{k-1}$ is flat, then one can stop at order $k$ in the prolongation process, and introduce at each point $p\in M$ a system of geodesic normal coordinates that contains the metric and it's second and higher derivatives up to the $k$th order (i.e., the Riemann tensor and its derivatives up to $k-2$-th as functions on $FM$).  Such a coordinate system 
\footnote{The system is not unique, but is unique up Lorentz transformation, and 
one can actually measure the size of the higher derivatives of $g$ by considering all the locally geodesic normal coordinates systems at $p$.\cite{Nash}}
 offers the best approximation of the spacetime geometry by $k$-th order tangent planes:  $g_{ab}(x^{c}) = \eta_{ab} + g_{ab, cd}x^{c}x^{d}+ \cdots$

Assuming that $M$ is normal hyperbolic spacetime (i.e., all points are regular), $c(P)$ can be computed \cite{Nomizu} in terms of scalar curvature invariants (as polynomial expressions in the curvature tensor and its first $(k-2)$-covariant derivatives regarded as functions on $FM$. In other words, the true observable $c$ provides in fact a local invariant classification of the gravitatinal metric \footnote{It only requires the knowledge of the curvature tensor and its covariant. derivatives up to sufficiently high order}

The motivation behind using coframes, as originally developed by Cartan\cite{Cartan, Chern},  was to solve the local equivalence problem, which is of central importance to general relativity.
Cartan scalars are a set of invariants defined by the Riemann tensor and its derivatives that will locally characterize a spacetime\cite{ExactSolutions}. 
($k\leq 10$ corresponds to the last derivative at which no functionally independent scalar on $FM$ arises). 

For spacetimes which have singular points, scalar invariants don't give a proper local characterization of spacetimes. \cite{Konkowski}

\begin{quote}
"...the metric may have parameters which are important
globally but do not appear in the Cartan scalars" and "The parameters cannot change the values of the Cartan scalars defined by the Riemann tensor and its derivatives at a point, and this directs attention to the possible global holonomy found by taking suitable closed curves..." \cite{MacCallum}
\end{quote}

At singular points the gravitational metric cannot be brought to the normal form, so singular points are not points of the smooth Lorentzian manifold $(M, g)$. In some cases, one can incorporate the singular points together with the regular spacetime points in some abstract set $\bar{M}$, equipped with a suitable topology that allows one to define statements such as 'close to the singularity' in a mathematically precise sense. In such situations, one may be able to study the spacetime geometry by adding additional structures on the Lorentzian $G$-structure $P$ representing the spacetime without the singular points.

A good example is given by the b-boundary of curvature singularities, where  $\bar M$ is formed by the projection on $M$ of the Cauchy completion of the Lorentzian structure of $P$ with respect to a Riemannian metric $\tilde  {g}$ (the b-boundary metric) constructed out of the soldering and connection 1-forms as follows:  
$$\tilde {g} (\xi ,\eta )= <\theta(\xi ),\theta(\eta )>_{V} + <\omega (\xi ),\omega (\eta )>_{\frak{g}}$$ for all for $\xi ,\eta \in TP$. 

One can compute \cite{Stahl} the scalar curvature form $\tilde R$ of the Riemannian geometry$(P,\tilde {g})$ in terms of the frame components of the Riemann tensor $R^{i}_{jkl}$ and its covriant derivatives of the Lorentzian structure $(M,g)$ is given by:
$$\displaystyle\tilde{R} = \frac{-n^{2}(n+3)}{2} - 1/4R^{i}_{jkl}R^{i}_{jkl} + R_{ii}$$
(Einstein summation over repeated indices)

It implies that an incomplete (in the b-metric $\tilde g$) endless curve $\gamma \in M$ has an enpoint $p$ on the bboundary $\bar{M}\setminus  M$, and its horizontal lift $\tilde \gamma$ has a finite b-length. The relation also suggests that scalar polynomial invariants together with scalar invariants constructued out of Cartan invariants are both to be used in order to analyse the concept of spacetime singularities. In \cite{Herrera} the authors analyse the concept of active gravitational mass for Reissner- Nordstrom spacetime in terms of scalar polynomial invariants and the Karlhede classification. In \cite{Henry} the author uses the Kretschmann scalar to find the amount of curvature of spacetime as a function of position near (and within) a Kerr-Newman black hole, which allows one to display the appearance of the black hole itself.

\section{DeRham cohomology classes for gravitational field}

Historically, differential forms proved to be one the most naturally sensitive objects to global aspects of manifolds. 

As suggested by Dirac, a more appropriate geometric model for electromagnetic theory (satisfactory from both classical and quantum point of view) is to interpret the 4-potential $A$ as a special sort of (horizontal) 1-form on a line bundle  $P\to M; S^{1})$, and the electromagnetic field $F$ as an element in the the second-integral cohomology group $H^{2}(M, \mathbb{Z})$, noticing that elements of $H^{2}(M, Z)$ are in 1:1 correspondence with the isomorphism classes of $S^{1}$-bundles over $M$

$\Omega (M)=(\oplus \Lambda ^{k}(M), \wedge , [,])$ is the DeRham cochain complex (graded commutative Lie algebra), and 
$H^{k}(M)=\{$ solutions of $d\omega =0$ modulo trivial solutions $d\sigma \}$ the DeRham cohomology groups.

We shall outline here the construction of new true observables as characteristic cohomology classes in the deRham groups of $M$. 

Let $(P\to M; G)$ be a Lorentzian $G$-structure on $M$ defining the gravitational field under consideration. The Lorentz group $G$ acts: on $P$ (on the right), on $V=\mathbb{R}^4$ ( on the left), on $\frak g$ (adjoint representation), and induces actions on multilinear forms: \\
on $L^{k}(V)$: for any $t\in L^{k}(V)$, $a\in G$, then\\
$\rho (a)t(v_{1},\cdots v_{k}) =t(a^{-1}v_{1},\cdots a^{-1}v_{k})$,\\
on $S^{k}(\frak {g}*)$ : for 
any $t\in S^{k}(\frak {g}*)$, $a\in G$, then\\
$\rho (a)t(v_{1},\cdots v_{k}) =t(Ad(a^{-1})v_{1},\cdots Ad(a^{-1})v_{k})$\\
on $\frak {g}^k$ : for any 
$t\in \frak {g}^k $, $a\in G$, 
then $\rho (a)t(v_{1},\cdots v_{k}) =at(a^{-1}v_{1},\cdots a^{-1}v_{k})$.\\

Let $\omega \in \Lambda ^{1}(P; g)^{G}$, $\theta \in\Lambda ^1_{\text{hor}}(P;V)^G$ $\tau \in\Lambda ^2_{\text{hor}}(P;V)^G$ and $\Omega \in\Lambda ^{2}(P;\frak{g})^G$ the connection, soldering, torsion and curvature form, respectively.  These forms are related via Cartan's structure equations: $\tau =d\theta +r_{*}(\omega) \wedge \theta  $ and $\Omega  =d_{\omega}\omega= d\omega  + 1/2[\omega, \omega] $, where the product of a $\frak{g}$-valued form $\alpha\in\Lambda ^{p}(P;\frak{g})$ with a $V$-value form $\lambda\in\Lambda^{q}(P;V)$ is given as the wedge product $\rho_{*}(\alpha)\wedge\lambda\in\Lambda ^{p+q}(P;V)$, and $\rho_{*}$ denotes the representation of the Lie algebra $g$ induced by the linear representation $\rho$.

The curvature form $\Omega \in\Lambda ^{2}(P;g)$ satisfies the Bianchi identity  $d_{\omega}\Omega =0$  and for $\omega $ Levi-Civita connection on $M$, $d_{\omega}\theta =0$.

One can compose a $V$-valued differential form $\psi_{i}\in\Lambda ^{p_i}(P;V)$ with a multilinear map
$f\in L^{k}(V)=(\bigotimes^{k} V)^*$ we get an ordinary differential form:

$ f^{\psi_1,\dots,\psi_k}:= f\circ (\psi_1\otimes_\wedge \dots \otimes_\wedge \psi_k )\in\Lambda ^{p_1+\dots+p_k}(P)$, where the product of vector-valued forms is defined as
$\otimes:\Lambda^{p}(M,V)\times\Lambda^{q}(M,W)\to\Lambda ^{p+q}(M,V\otimes W)$
(with the exception that real multiplication is replaced with the tensor product $\otimes $)

If $\psi_i\in \Lambda ^{p_i}_{\text{hor}}(P;V)^G$ and 
$f\in L^{k}(V)^G$ are $G$-invariant and horizontal then 
$f^{\psi_1,\dots,\psi_k}$ is $G$-invariant and horizontal, and so it represents the pullback of a ordinary form on $M$.

\subsection{Algebra homomorphisms}

The Chern-Weil homomorphism is a basic construction relating the curvature to the deRham cohomology groups of $M$ i.e., the geometry and the topology.

The algebra $R(g^{*})^{G}$ of the $\mathbb R$-valued homogeneous polynomial on $g^*$ of degree $k$ is isomorphic with  $(Sym^{k}(g^{*}))^{G}$. 

The Chern-Weil homomorphism is a homomorphism of algebras from $(Sym^{k}(g^{*}))^{G}$  to the deRham cohomology algebra $H^*(M)$ defined as follows: for any $f\in (Sym^{k}(g^{*}))^{G}$ one can define a 2k-form $f^{\Omega }$ on $P$ given by:

$f(\Omega)(X_1,\dots,X_{2k})=\frac{1}{(2k)!}\sum_{\sigma\in\mathrm S_{2k}}\epsilon_\sigma f(\Omega(X_{\sigma(1)},X_{\sigma(2)}),\dots,\Omega(X_{\sigma(2k-1)},X_{\sigma(2k)}))$
 
where $\epsilon_\sigma$ is the sign of the permutation $\sigma $ in the symmetric group on $2k$ numbers $\mathrm S_{2k}$. The deRham cohomology class of $f^{\Omega }$ depends only upon the G-structure $P$

One can define a homomorphism of algebras from $\Lambda (V^{*})^{G}$  to the deRham cohomology algebra $H^*(M)$ by $f\longmapsto f^{\theta}= f(\theta ,\dots,\theta )$
constructed using the soldering form. Let $\omega$ be the 1-form connection on $P$ corresponding to the Levi-Civita connection on $(M,g)$. Then $d\tau =d_{\omega }\theta =0$ which implies $f^\theta$ is closed. The deRham cohomology class of $f^{\theta}$ is independent of the choice of the torsion-free connection form on $P$, so it only depends upon the Lorentzian $G$-structure $P$. 

Let $(S(\frak{g}^*))^{G}\otimes (\Lambda (V^{*}))^G$ the associative algebra of the $G$-invariant elements. There is a homomorphism of algebras from $A(\frak{g},V)^{G}$   to the deRham cohomology algebra $H^*(M)$ given by $\Upsilon (f)= f^{\Omega ,\theta} = f(\Omega,\dots,\Omega, \theta ,\dots,\theta)$. (see\cite{Michor})
 
The deRham cohomology class of $f^{\theta,\Omega}$ is independent of the choice of the torsion-free connection 1-form on $P$, so it depends only upon the Lorentzian $G$-structure $P$. For $f\in (S^{p}(\frak{g}^*))^{G}\otimes (\Lambda^{q} (V^{*}))^G$ denote $c_{f}(P)$ the cohomology class of $f^{\theta,\Omega}$ in $H^{2k+l}(M)$. $c_{f}(P)$ is an invariant of the $G$-structure $P$, and defines a true observables of the associated gravitational field.

{\em Examples:}

(1) The cohomology class of $\Upsilon (Pf) $ the Pfaffian (defined for $A\in\frak{g}$ by 
$\displaystyle\mathrm{Pf}(A) = \frac{1}{2^{n} n!}\sum_{\sigma\in S_{2n}}\mathrm{sgn}(\sigma)\prod_{i=1}^{n}a_{\sigma(2i-1),\sigma(2i)}$)
is the Euler class of the Lorentzian structure $P$.\\

(2) The cohomology class of Pontrjagin polynomials $P_{k/2}$ ($k$ even) under the $\Upsilon$ map are the Pontrjagin classes of the $G$-structure $P$.\\

(3) The Lie algebra $so(1,3)\in\Lambda^{2}V$ (consists of skew-symmetric matrices). The $G$-equivariant projection of $\Lambda^{2}V$ to $so(1,3)$ along a suplement of the $so(1,3)$ in $\Lambda^{2}V $ defines a $G$-invariant element $f\in so(1,3)^{*}\otimes\Lambda ^{2}(V^{*})$ ( where $Sym^{1}(g^{*})= so(1,3)^{*}$). Its image under the $\Upsilon$- homomorphism defines a 4-form $f^{\Omega, \theta,\theta}$ on $M$ has the local expressions  $f_{abc}^{d}R_{def}$ in an  orthonotmal basis $(e_{a})$. (Where the local expressions for $f$ and the curvature tensor are $f_{abcd}$ and $\Omega^{a}_{b} = \Omega(e_{a}, e_{b})= R_{bef}^{a}e_{e}, e_{f}$ respectively).\cite{Michor}

Other scalar polynomial invariants can be obtained by choosing an appropriate linear combinations of $G$-invariant polynomials, and integrating the forms on $M$. 

Example of polynomial invariants, are the curvature invariants:
the Kretschmann scalar  $K_1 = R_{abcd} \, R^{abcd}$, the Chern-Pontryagin scalar $K_2 = {{}^\star R}_{abcd} \, R^{abcd}$
and the Euler scalar $K_3 = {{}^\star R^\star}_{abcd} \, R^{abcd}$. In total there are 14 algebraically independent scalar invariants\footnote{local invariants that don't satisfy a polynomial relation}  constructable from the metric and curvature. Any attempt to find a complete set of scalar invariants using the 14 algebraically independent scalars failed(see e.g., \cite{ExactSolutions})
 
Although scalar polynomial invariants are insufficient in providing a local characterization of all spacetimes uniquely\cite{Malcolm}, together with Cartan scalars they are useful in studying spacetime with singularities, and the geometry of black holes. In fact, in \cite{Henry} a meaningful picture of a black hole has been obtained by plotting the Kretschmann scalar polynomial.

\section{Some conclusions. Open problems}

In order to get some insight into the global structure of the spacetime geometry, in this paper I investigated certain aspects of general covariance. I constructed a set of true geometric observables of the classical gravitational field as global spacetime diffeomorphism invariant cohomology classes. They represent global characteristics of the physical field, are linked to the topology of the spacetime, and define uniquely a local characterization of the spacetime. 

The problem of finding a complete set of data out of geometric observables is related to the fact that $4$-dimensional manifolds cannot be effectively classified: given two spacetime manifolds, there is no algorithm for determining if they are isomorphic (or diffeomorphic). This may suggest that the spatio-temporal structure of our universe might be theoretically underdetermined.
 One way of dealing with inherent uncertainty due to limited information of the global characterisation of the spacetime is to use statistical analysis to try to quantify the uncertainty in the missing data.\\

Open Problem: Using the language of $G$-structures one could gain some insight into how to study the geometry of spacetimes with curvature singular points (e.g., the bboundary) by trying to find some additional (compatible) geometric structures on the Lorentzian structure $P$, and then compute the local and global invariants as shown in this paper.

\end{document}